\documentclass[12pt]{article}

\begin{document}

\begin{center}
{\LARGE{On evidence for negative energies and masses in the Dirac equation through a unitary time-reversal operator}}
\end{center}

\begin{center}
N. Debergh $^1$, J.-P. Petit $^2$ and G. D'Agostini $^3$
\vspace{5mm}\\
{\footnotesize{$^1$ Department of agronomy, Haute Ecole Charlemagne, 3, rue Saint-Victor, 4500 Huy, Belgium, nathalie.debergh@hech.be\vspace{2mm}\\
$^2$ jp.petit@mailaps.org\vspace{2mm}\\
$^3$ dagostinigilles@mailaps.org}}
\end{center}

\begin{abstract}
We review the standards of relativistic quantum mechanics such as the Dirac equation under the concept of negative masses. We show that negative energies are acceptable provided the masses are simultaneously negative. Negative energy and mass anti-fermions are obtained from positive energy and mass fermions through a unitary PT transformation.
\end{abstract}

\section{Introduction}
Until recently, the concept of negative masses was relegated to the field of “exotic”, and even far-fetched, ideas. 
\par

Proposed by Hermann Bondi \cite{Bondi} and revisited by William B. Bonnor \cite{Bonnor} in the context of general relativity, negative masses have been considered by a few physicists only due to the fact that they have never been detected in laboratories. Let us mention among these avant-garde physicists, M.A. Markov who, from a previously proposed second-order equation for spin ${\frac{1}{2}}$ particles \cite{Markov1}, has put in evidence two first-order equations, one with a "plus"' sign and the other one with a "minus" sign multiplying the mass \cite{Markov2}. Nevertheless, for the reason of being qualified as "non-physical"and because they could imply a very unpleasant “runaway effect”, negative masses have been rejected by the majority of the scientific community for many years. 
\par

However, minds are evolving nowadays : the possibility of negative masses is seriously reconsidered. As examples, let us mention that negative mass solutions have been considered within a Schwarzschild–de Sitter geometry \cite{Belletete} or in various aspects of physics \cite{Hammond}, going from wormholes to strings. In \cite{Mbarek}, it was shown that negative mass can produce a Schwarzschild solution without violating the energy condition. The Dirac equation has also been considered with the two possibilities of signs for the mass and it was shown \cite{Dvoeglazov1} that positive as well as negative energies can arise.
\par
Besides these theoretical contributions, a negative effective inertial mass has been experimentally realized by cooling rubidium atoms with lasers as stated in \cite{Khamehchi}.
\par

A further reason for this renewed interest in negative masses is the following. We have mentioned the “runaway effect” as one of the main causes for leaving these masses away. What is it from ? As well known, general relativity stands that Universe is a manifold equipped with a metric $g_{\mu \nu}$ according to the Einstein equation
$$
R_{\mu \nu}-\frac{1}{2}  R g_{\mu \nu}= \chi T_{\mu \nu} 
$$ 
The matter energy tensor $T_{\mu \nu}$ is the source of the gravitational field. 
\par
The paradox comes from the fact that it is arbitrarily assumed that particles of positive or negative masses immersed in the same gravitational field, react in the same way. Indeed, as there is only one metric, the effect on each particle from that field created by a positive or negative (in that case $T\rightarrow -T$) mass is the following : 
\begin{itemize}
	\item Positive mass attracts both other positive masses and negative masses 
	\item Negative mass repels both other negative masses and positive masses 
\end{itemize}
This means that the particle of negative mass pursues, in a uniformly accelerated movement with a  conserved kinetic energy, the particle of positive mass. Indeed, due to their respective signs, the kinetic energy of the positive mass is  simultaneously, while remaining in the same ratio, compensated by that of the positive mass irrespective of the increase of the speed.
\par

Besides the fact that this “runaway effect” violates the action-reaction principle, it also disagrees with experimental data \cite{Riess}. 
\par
If we deviate from the assumption that positive and negative masses react in the same way, we need two metrics and a set of two field equations, generalizing the Einstein one. A model ( the Janus cosmological model or JCM) with such two coupled field equations has been proposed \cite{Petit1}. The subsequent Newtonian approximation provides the following corresponding interaction rules : 
\begin{itemize}
	\item Positive mass attracts positive masses and repels negative masses
	\item Negative mass attracts negative masses and repels positive masses
	\end{itemize}
The JCM perfectly fits available observational data \cite{Petit2} and restores the action-reaction as well as the equivalence principles. This certainly gives substance to the study of negative masses.
\par

Despite of the growing interest in negative masses, there were very few papers devoted to the subject within the quantum mechanical paradigm. Let us cite \cite{Yu} where the concept of negative mass is used to solve Schrödinger equations.
\par 

The purpose of this article is to initiate a study of the impact of negative masses on relativistic quantum mechanics by taking a new look at its standards such as the Dirac equation (for a review of these standards, see, for instance, Section 2.3 of \cite{Ryder}). 
\par
The reason to do so is that relativistic quantum mechanics naturally puts in evidence negative energies having an unpleasant consequence which can be avoided if masses are negative, too. Indeed, one of the arguments for having given up the Klein-Gordon equation is the fact that the density of probability is, up to the square of a normalization factor, the ratio $\frac{E}{m}$. It is obvious to see that this probability is positive if energies and masses have the same sign. So, negative energies can be restored if masses are negative, too.
\par

The contents is distributed as follows. In Section 2, we show that if the time-reversal operator is chosen as unitary, masses become negative (while an anti-unitary one leads to positive masses). In Section 3, we realize this unitary operator within the Dirac formalism. States corresponding to antifermions are then associated with negative masses and energies. We conclude in Section 4.
\par
 
We work in the following in the standard units. Greek letters run from 0 to 3 while Latin indices are equal to 1, 2, 3.

\section{Unitarity of the time-reversal operator}
First, let us recall that a unitary operator U is defined through
\begin{equation}
< U \psi | U \phi> = <\psi | \phi> 
\label{(1)}
\end{equation}
for all states $|\psi>$ and $|\phi>$ of the Hilbert space H. It is also a linear operator 
\begin{equation}
U (a |\psi> + b |\phi>) = a \; U |\psi> + b \; U |\phi> \; , \; a,b \in C
\label{(2)}
\end{equation}
An operator A is anti-unitary if it obeys
\begin{equation}
< A \psi | A \phi> = <\psi | \phi>^{\ast}  
\label{(3)}
\end{equation}
where the symbol $\ast$ denotes the complex conjugation. \\
Furthermore, it fills a similar relation to Eq. (2) but where the second member has been conjugated, meaning that A is anti-linear.
\par
Usually, one can see any anti-unitary operator as the product of a unitary one and the complex conjugation (i.e. $ i \rightarrow -i$).
\par 
Such operators are crucial in quantum mechanics due to the fact that E.P. Wigner \cite{Wigner} showed that a symmetry operator S of a Hamiltonian is necessarily  a unitary one or an anti-unitary one.
\par
In what concerns the discrete symmetries, it is usually understood that the parity operator P such that
\begin{equation}
\vec x \rightarrow - \vec x \; , \; \vec p \rightarrow - \vec p \; , \; t \rightarrow t
\label{(4)}
\end{equation}
is a unitary one while the time-reversal operator T implying
\begin{equation}
\vec x \rightarrow  \vec x \; , \; t \rightarrow -t
\label{(5)}
\end{equation}
is anti-unitary (in which case, $\vec p \rightarrow - \vec p$)
\par
The reasons for such a belief are the following ones. First, if we want the energies to stay positive, we have to add a complex conjugation to the reversal of time because of the quantization
\begin{equation}
E \leftrightarrow i \frac{\partial}{\partial t}
\label{(6)}
\end{equation} 
Second, this is a way to remain the fundamental relation of quantum mechanics
\begin{equation}
[ x_j , p_k ] = i \delta_{jk}
\label{(7)}
\end{equation}
invariant under the transformation PT as its action can be summarized by
\begin{equation}
\vec x \rightarrow - \vec x \; , \; \vec p \rightarrow \vec p \; , \; E \rightarrow E \; ,i \rightarrow -i
\label{(8)}
\end{equation}
These developments suggest that both energies and masses are positive. These signs are thus related to an anti-unitary PT.\par
We argue that they can be understood in a different way. \par
Indeed, Eq. (6) is coherent with T being unitary if we allow the energy to be negative while Eq. (7) is still true under a unitary PT in  which case Eq. (8) has to be replaced by
\begin{equation}
\vec x \rightarrow - \vec x \; , \; \vec p \rightarrow -\vec p \; , \; E \rightarrow -E \; , i \rightarrow i
\label{(9)}
\end{equation}
The fact that the sign of the energy changes  has consequences on masses. This can be seen, for example, through the Einstein relation $E = m$.
\par
Thus we have two options. The usual one, dealing with an anti-unitary PT symmetry and ensuring the energies and masses to stay positive together, and the one we propose, that is a unitary PT symmetry (P and T are both unitary operators) which allows energies and masses to simultaneously be negative.
\par
Note that this second path is in perfect agreement with what we know about the Lorentz group. Indeed, being defined as the group of matrices $\Lambda$ satisfying 
$$
\Lambda^T  \; g \; \Lambda = g, \; g = diag (1, -1, -1, -1)
$$
where $\Lambda^T$ is the transpose of $\Lambda$,the entire Lorentz group is composed of four components \cite{Rao} 
\begin{itemize}
\item $L_+^{\uparrow}$
\item $L_-^{\uparrow}$ = P $L_+^{\uparrow}$
\item $L_-^{\downarrow}$ = T $L_+^{\uparrow}$
\item $L_+^{\downarrow}$ = P T $L_+^{\uparrow}$
\end{itemize}
where the subscript $\pm$ denotes the signs of the determinant of $\Lambda$ while the superscript refers to the signs of the first element of this matrix. \\
As is well known, there are transformations that maps one component into another. These transformations are nothing but the parity and time-reversal operators realized here through
\begin{equation}
P = diag (1, -1, -1, -1) \; , \; T = diag (-1, 1, 1, 1)
\label{(10)}
\end{equation}
Let us take a while here to notice that both P and T being realized through such matrices are linear and unitary.\\
Only a few components do form a group, i.e. $L_+^{\uparrow}$ (named the restricted Lorentz group ; it is the one considered by physicists exclusively concerned with positive masses and energies ), $L_+^{\uparrow} \cup L_-^{\uparrow}$ (named the orthochronous Lorentz group), $L_+^{\uparrow} \cup L_+^{\downarrow}$ (named the proper Lorentz group ), $L_+^{\uparrow} \cup L_-^{\downarrow}$ and the set of four components referred to as the entire Lorentz group.\\
Only one of these options, the proper Lorentz group, is recognized as a known Lie group, namely $SO(1,3)$
$$ 
L_+^{\uparrow} \; \cup \; P T L_+^{\uparrow} \; \Rightarrow \; SO(1,3)
$$

If we now add to these transformations the four abelian translations and their PT version, we obtain two components generating a subgroup of the entire Poincaré group. One thing well-known about this group is that it admits two Casimir operators, the first one having the square of the mass as eigenvalues. The only requirement on these eigenvalues is their reality. The mass itself can thus be positive or negative. Besides, the meaning of the second Casimir operator constructed from the Pauli-Lubanski 4-vector is clear when limited to isotropy groups. In this case, one can choose a particular time-like momentum 4-vector, say
\begin{equation}
\left(
\begin{array}{c}
m \\
0 \\
0\\
0
\end{array}
\right)
\label{(11)}
\end{equation}
and see that the second Casimir operator is related, up to the square of the mass (again…), to the spin of the particles. These are very well-known facts. 
\par
Now, let us have a look on the 4-vector  (11). If the mass is positive (respectively negative), it belongs to the future (respectively past) light cone. An orthochronous transformation is realized through a matrix whose first element is bigger than one (it is the "$\uparrow$" symbol). So, when applied to a 4-vector like Eq. (11), it gives rise to another time-like 4-vector within the same part of the cone light : if the mass is positive and the 4-vector belongs to the future part, its image will also stand in the future part. If the mass is negative, the 4-vector lives in the past part and stays in this part after an orthochronous transformation. Only antichronous transformations (the "$\downarrow$" symbol) can join the two parts. If the mass is positive at the start and the four-vector oriented from past to future, it will change its direction but also the sign of the associated mass. This is due to the fact that the antichronous transformations result from the application of a unitary PT (= $- diag(1,1,1,1)$) on the orthochronous ones, changing the sign of the first component of 4-vectors. And  if the mass becomes negative, so is the energy through the Einstein relation  $E=m$  for a relativistic particle at rest. 
\section{Realization of the unitary PT symmetry for the Dirac equation}
Let us now see the impact of a unitary PT transformation on the Dirac equation. 
\\
For recall, the covariant Dirac equation is written as \cite{Dirac1}
\begin{equation}
(\gamma^{\mu} p_{\mu} - m) \psi = 0
\label{(12)}
\end{equation}
with the Dirac matrices satisfying
\begin{equation}
\{ \gamma^{\mu} , \gamma^{\nu} \} = 2 g^{\mu \nu }\; , \; g = diag(1,-1,-1,-1)
\label{(13)}
\end{equation}
We realize these matrices through the so-called Dirac representation i.e. 
\begin{equation}
\vec \gamma = \left( 
\begin{array}{cc}
0 & \vec \sigma \\
- \vec \sigma & 0
\end{array}
\right) \; , \;
\gamma^0 = \left( 
\begin{array}{cc}
I & 0 \\
0 & -I
\end{array}
\right)
\label{(14)}
\end{equation}
where the $\vec \sigma$ are the Pauli matrices.
\\
The usual unitary parity operator P is realized through
\begin{equation}
\psi (t, \vec x) \rightarrow \gamma^0 \psi (t, -\vec x)
\label{(15)}
\end{equation}
In what concerns a unitary time-reversal, we have to ask for 
\begin{equation}
T_U (i \gamma^0 \partial_t + i \gamma^j \partial_j -m)T_U^{-1} T_U \psi(x) = (-i \gamma^0 \partial_t + i \gamma^j \partial_j -m) \psi'(x')=0
\label{(16)}
\end{equation}
with
\begin{equation}
x=(t, \vec x) \rightarrow x'=(-t, \vec x)
\label{(17)}
\end{equation}
In other words, we have to search for a matrix $T_U$ anticommuting with $\gamma^0$ and commuting with each of the $\gamma^j$'s. This matrix is $\gamma^1 \gamma^2 \gamma^3$.
Up to a phase, the action of the time reversal as a unitary operator is thus marked by
\begin{equation}
\psi (t, \vec x) \rightarrow \gamma^1 \gamma^2 \gamma^3 \psi (-t, \vec x)
\label{(18)}
\end{equation}
Combined with the parity operator (15), we are led to the following unitary PT transformation 
\begin{equation}
\psi (t, \vec x) \rightarrow i \gamma^0 \gamma^1 \gamma^2 \gamma^3 \psi (-t, -\vec x) = \gamma^5 \psi (-t, -\vec x)
\label{(19)}
\end{equation}
The PT-symmetry within the Dirac formalism is not just only related to opposite space-time coordinates, it is also subtended by the $\gamma^5$ matrix. \par
We are used to interpret this matrix as the one realizing the chirality operator. But it is not just so. Indeed, the $\gamma^5$ matrix anticommutes with each of the $\gamma^{\mu}$'s. Acting on the Dirac equation (12), this implies that either the sign of each of the $p_{\mu}$'s has to change (which is expected in a unitary PT symmetry) or, in an equivalent way, the mass has to become negative. Such developments can also be found in the quantum field theory context \cite{Dvoeglazov2} where it was quoted that the $\gamma^5$ chiral transformation is associated to the fact that the sign of the mass term in the Lagrangian is reversed. 
Let us also mention as an aside that the $\gamma^5$ matrix has also been associated to the charge operator \cite{BarutZiino}. More precisely, in \cite{BarutZiino}, it was required that the Dirac parity operator should always be of the form (15) whether it is defined in the fermion or antifermion wavefunction space. And this had consequences that the fermion field and the charge-conjugate (related to $\gamma^5$ ) antifermion field obey opposite-mass Dirac equations. However, despite of the fact that this approach is close to ours, it is different in the sense that we are not only in the QFT field but mostly because we are concerned here with negative energies.\par Together with previous remarks, this fact tends to prove that the matrix $\gamma^5$ has not yet revealed all its wealthes...\par
Anyway, the matrix  $\gamma^5$ is essentially considered here as the one for what we call a M-symmetry :$ \; m \rightarrow -m$. Such a symmetry is naturally connected to a unitary PT-transformation.\par
Let us now analyse the impact of such considerations on the four solutions of the Dirac equation.\\
As well known, the free Dirac equation (12) has positive as well as negative energies $p_0 = \pm E_p = \pm \sqrt{p^2 + m^2}$. The four solutions  consist in two solutions with positive energies $E= E_p  $ (spin up and spin down, respectively, according to the eigenvalues of the third component of the spin vector $S_3 = \frac{1}{2}\; diag(1,-1,1,-1)$) which write in the p-representation (in which $p_{\mu}$ now stand for eigenvalues)
\begin{equation}
\psi_1^+ = \sqrt{\frac{E+m}{2E}}\; e^{i (\vec p . \vec x - E t)} \left( 
\begin{array}{c}
1 \\
0 \\
\frac{p3}{E+m} \\
\frac{p1+ip2}{E+m}
\end{array}
\right) \; , \; \psi_2^+ = \sqrt{\frac{E+m}{2E}}\; e^{i (\vec p . \vec x - E t)} \left( 
\begin{array}{c}
0 \\
1 \\
\frac{p1-ip2}{E+m} \\
-\frac{p3}{E+m}
\end{array}
\right)
\label{(20)}
\end{equation}
and two solutions with negative energies $E=- E_p$ (spin up and spin down, respectively)
$$
\psi_1^- = \sqrt{\frac{E-m}{2E}}\; e^{i (\vec p . \vec x - E t)} \left( 
\begin{array}{c}
\frac{p3}{E-m} \\
\frac{p1+ip2}{E-m} \\
1 \\
0
\end{array}
\right) \; , \; \psi_2^- = \sqrt{\frac{E-m}{2E}}\; e^{i (\vec p . \vec x - E t)} \left( 
\begin{array}{c}
\frac{p1-ip2}{E-m} \\
-\frac{p3}{E-m} \\
0 \\
1
\end{array}
\right)
$$
which are replaced by two solutions with positive energies $E= E_p$ (spin down and spin up, respectively)
\begin{equation}
\chi_1^+ = \sqrt{\frac{E+m}{2E}}\; e^{-i (\vec p . \vec x - E t)} \left( 
\begin{array}{c}
\frac{p3}{E+m} \\
\frac{p1+ip2}{E+m} \\
1 \\
0
\end{array}
\right) \; , \; \chi_2^+ = \sqrt{\frac{E+m}{2E}}\; e^{-i (\vec p . \vec x - E t)} \left( 
\begin{array}{c}
\frac{p1-ip2}{E+m} \\
-\frac{p3}{E+m} \\
0 \\
1
\end{array}
\right)
\label{(21)}
\end{equation}
by the transformation
\begin{equation}
\vec p \rightarrow - \vec p \; , \; E \rightarrow - E
\label{(22)}
\end{equation}
Actually, the states (21) are solutions of a slightly different Dirac equation
$$
(\gamma^{\mu} p_{\mu} + m) \chi = 0
$$
The changes (22) are attributed to Feynman \cite{Feynman} and Stueckelberg \cite{Stueckelberg}, independently. They were motivated by two reasons : first, keep positive energies only and second, give a physical interpretation to these states. Indeed, in case of an electromagnetic interaction, the corresponding negative energy solutions don’t give back a meaningful Pauli equation at the non-relativistic limit while the new positive energy ones (21) do with an electric charge having been replaced by its opposite. This is what have led physicists to interpret these new solutions as the ones associated with antimatter. 
\par
Let us now analyse these solutions (20) and (21) by keeping in mind that both energies and masses can be simultaneously negative. \\
Besides the bi-spinors, we notice that the link matter-antimatter is related to the following  transformation of the exponential
\begin{equation}
e^{i (\vec p . \vec x - E t)} \rightarrow e^{-i (\vec p . \vec x - E t)}
\label{(23)}
\end{equation}
which can be performed in three different ways.\\ 
First, the complex conjugation  : this is the path followed by Dirac \cite{Dirac2}. Its anti-unitary charge conjugation operator is just the complex conjugation in the Majorana representation but writes
\begin{equation}
C = i \gamma^2 \; {\bf{C}}
\label{(24)}
\end{equation}
in representation (14). \\
It is easy to verify that 
\begin{equation}
C \; \psi_1^+ (x) = \chi_2^+ (x) \; , \; C \; \psi_2^+ (x) = -\chi_1^+ (x) 
\label{(25)}
\end{equation}
The anti-unitary charge conjugation operator thus transforms positive energy and mass fermions into positive energy and mass anti-fermions. The mix between $\psi_1-\chi_2$, $\psi_2-\chi_1$ in Eqs (25) is related to the fact that the charge conjugation operator is such that
\begin{equation}
\vec x \rightarrow \vec x \; , \; \vec p \rightarrow - \vec p \; \Rightarrow \vec L \rightarrow - \vec L
\label{(26)}
\end{equation}
Consequently, the sign of the spin vector has also to change in order to stay coherent. This implies that $C$ acts on spin up (resp. down) $\psi$-states to give spin up (resp. spin down) $\chi$-states. Let us insist on the fact that these developments concern the first quantization field only. In quantum field theory, the charge conjugation operator is somewhat different and appears, in particular, to be unitary. \vspace{2mm}\\
Second, the transformation (22) : it is what Feynman (and Stueckelberg) proposed. This is associated to an anti-unitary PT transformation in the quantum field theory context. However, when we limit ourselves to the first quantization as we do here, such a transformation cannot relate $\psi$-states to $\chi$-states. Indeed, the anti-unitary reversal-time transformation would be the one ensuring a relation similar to (16)
\begin{equation}
T_{AU} (i \gamma^0 \partial_t + i \gamma^j \partial_j -m)T_{AU}^{-1} T_{AU} \psi(x) = (i \gamma^0 \partial_t - i (\gamma^j)^{\ast} \partial_j -m) \psi'(x')=0
\label{(27)}
\end{equation}
meaning that $T_{AU}$ has now to commute with $\gamma^0$ and $\gamma^2$ (the only matrix among the $\gamma^j$'s to have imaginary elements) and anticommute with $\gamma^1$ and $\gamma^3$. In other words, the action of $T_{AU}$ would be such that
\begin{equation}
\psi (t, \vec x) \rightarrow i \gamma^1 \gamma^3 \psi (-t,  \vec x)^{\ast}
\label{(28)}
\end{equation}
for the Dirac equation we are concerned with. Combined with the parity (15), it gives rise to the following anti-unitary PT 
\begin{equation}
\psi (t, \vec x) \rightarrow i \gamma^0 \gamma^1 \gamma^3 \psi (-t, - \vec x)^{\ast}
\label{(29)}
\end{equation}
which it is easy to see that it simply connects  positive energy and mass fermions to positive energy and mass fermions
\begin{equation}
\gamma^0 \gamma^1 \gamma^3 \psi_1^+(-x)^{\ast} = - \psi_2^+ (x) \; , \; \gamma^0 \gamma^1 \gamma^3 \psi_2^+(-x)^{\ast} =  \psi_1^+ (x)
\label{(30)}
\end{equation}
This is in complete agreement with what we quoted in Section 2 : only unitary PT can rely the two parts of the light cone. 
\vspace{2mm}\\
So, we come to the third proposal : a unitary PT as given in Eq. (19) which, let us insist on that point, does not only perform the change (23) by replacing the coordinates by their opposites but also implies the M-symmetry $m \rightarrow -m$.\\
The explanation is obvious when considering the Dirac operator 
\begin{equation}
\left(
\begin{array}{cc}
E-m & - \vec \sigma . \vec p \\
\vec \sigma . \vec p & -E-m 
\end{array}
\right)
\label{(30)}
\end{equation}
acting on bi-spinors to give 0.
There are two options only : either you perform the changes (22) and leave in particular the negative energies to concentrate on positive ones
$$
\left(
\begin{array}{cc}
-E-m &  \vec \sigma . \vec p \\
- \vec \sigma . \vec p & E-m 
\end{array}
\right)
$$
or you choose to keep the negative energies but then, you are obliged to consider, simultaneously, negative masses
$$
\left(
\begin{array}{cc}
E+m & - \vec \sigma . \vec p \\
\vec \sigma . \vec p & -E+m 
\end{array}
\right)
$$ 
So, this third solution, based on a unitary PT, also leads to the solutions (21)
\begin{equation}
\gamma^5 \psi_1^+ (-x) = \chi_1^+ (x) \; , \; \gamma^5 \psi_2^+ (-x) = \chi_2^+ (x)
\label{(32)}
\end{equation}
as it can be easily verified, with the difference that the anti-fermions (21) now have negative energies and masses. \\
We insist on that fact : the $\chi$-states as in Eq. (21) remain the same but are interpreted in a different way. Under the unitary PT transformation, the following changes arise
$$
E \rightarrow -E \; , \; m \rightarrow -m \; ,\; x_{\mu} \rightarrow - x_{\mu} \; , \; p_{\mu} \rightarrow - p_{\mu}
$$
but they leave the different elements of the $\chi$-states invariant whether it is the normalization factor 
$$
\sqrt{\frac{E+m}{2E}}
$$
the plane wave part 
$$
e^{-i (\vec p . \vec x - E t)}
$$
or the fractions inside the bi-spinors because the signs of both numerators and denominators simultaneously change.\\
Another point is the fact that, in Eq.(32), by opposition to Eq.(25), the unitary PT connects the sectors $\psi_1-\chi_1$, $\psi_2-\chi_2$. This is due to the transformations of such an operator on position, momentum and angular momentum operators
\begin{equation}
\vec x \rightarrow \vec x \; , \; \vec p \rightarrow \vec p \; \Rightarrow \vec L \rightarrow \vec L
\label{(32)}
\end{equation}
(compare with Eq.(26)) implying the conservation of the spin vector. This implies that unitary $PT$ acts on spin up (resp. spin down) $\psi$-states to give spin up (resp. spin down) $\chi$-states. 
\section{Conclusion}
We have shown that negative masses can take place in relativistic quantum mechanics up to the condition of being related to negative energies. The unitary PT transformation we propose acts on states associated with positive energy and mass fermions to give states associated with negative energy and mass anti-fermions. This is an alternative to the anti-unitary charge conjugation operator leading to the same states but then subtended by positive energy and mass anti-fermions.\\
We argue, in complete agreement with \cite{Petit1}, that there are two types of antimatter. The "classical" one, created in laboratories and associated with an anti-unitary PT, and the primordial antimatter reached by the unitary PT, composed of the same elements, this time with a negative mass and energy. \\
It is this antimatter of negative mass (what we call negative antimatter) which is responsible for the acceleration of cosmic expansion, for the confinement of positive mass objects, the spiral structure, strong gravitational lens effects and so on. \\
Is there any place for a (unitary or anti-unitary) charge conjugation  operator leading to this negative antimatter ? This question remains open at this stage as well as the extension of our results to the quantum field theory context.

\end{document}